\documentclass[aps,prd,%multicol,twocolumn, 
preprintnumbers,groupedaddress]{revtex4}
\usepackage{epsbox}
\usepackage{amsmath}
\usepackage{bm}
\usepackage[dvips]{graphicx}

\begin{document}
\preprint{\vtop{
{\hbox{YITP-15-100}\vskip-0pt
%                 \hbox{KANAZAWA-??-??} \vskip-0pt
%                 \hbox{hep-ph/07????} 
}
}
}

\date{\today}

\title{ 
Decays of Charmed Vector Mesons \\ 
--- $\bm{\eta\pi^0}$ mixing as an origin of isospin non-conservation ---
}

\author{
Kunihiko Terasaki   %authors' name%
}
\affiliation{
Yukawa Institute for Theoretical Physics, Kyoto University,
Kyoto 606-8502, Japan 
%\\ Institute for Theoretical Physics, Kanazawa University, 
%Kanazawa 920-1192, Japan
}

\begin{abstract}
{
Pion emitting decays and radiative ones of $D^{*+,0}$ and $D_s^{*+}$ are studied. 
As the result, the full-width of $D^{*0}$ is predicted, by assuming the isospin 
symmetry. 
In addition, the isospin non-conserving $D_s^{*+}\rightarrow D_s^+\pi^0$ decay is 
investigated under the assumption that it proceeds through the $\eta\pi^0$ mixing. 
}               
\end{abstract}

\maketitle

It is seen in experiments~\cite{PDG14} that the charm strange vector meson 
$D_s^{*+}$ decays dominantly into $D_s^+\gamma$, while its isospin non-conserving 
$D_s^+\pi^0$ decay is much weaker. 
It is qualitatively understood by a hierarchy of hadron interactions, 
%%%%%%%%%%%%%%%%%%%%%%%%%%%%%%%%%%%%%%%%%%%%%%%
$|${\it isospin\,\, conserving ones} $\sim O(1)|\gg$ 
$|${\it electromagnetic ones} $\sim O(\sqrt{\alpha})|\gg$
$|${\it isospin non-conserving ones} $\sim O(\alpha)|$~\cite{Dalitz},   
%%%%%%%%%%%%%%%%%%%%%%%%%%%%%%%%%%%%%%%%%%%%%%%%%%%%%%%%%%%%%%%%%%%%%%%%
where $\alpha$ is the fine structure constant.  
In reality, it has been observed that this hierarchy plays an important 
role in decays of charm strange scalar $D_{s0}^+(2317)$~\cite{HT-isospin}. 
In addition, the measured width of $D^{*+}$ meson has been greatly improved 
recently~\cite{Babar-D^*}. 
Therefore, we study decays of charmed vector mesons $D^{*+,0}$ and $D_s^{*+}$ 
to see numerically a role of the above hierarchy, in particular, a role of the 
$\eta\pi^0$ mixing as an origin of isospin non-conservation. 

Using the updated width and branching fractions  
%%%%%%%%%%%%%%%%%%%%%%%%%%%%%%%%%%%%%%%%%%%%%%%%%%%%%%%%%%%%%%%%%%%%%%%%
\begin{eqnarray}
&& \hspace{10mm}
(\Gamma_{D^{*+}})_{\rm exp}\hspace{8mm} = 83.4\pm 1.8\,\, {\rm keV}, \,\,
Br(D^{*+}\rightarrow D^0\pi^+)_{\rm exp} = 67.7 \pm 0.5\,\, \% , 
\nonumber \\  
&& 
Br(D^{*+}\rightarrow D^+\pi^0)_{\rm exp} = 30.7 \pm 0.5\,\, \%, \quad\,
Br(D^{*+}\rightarrow D^+\gamma)_{\rm exp}\,\,\, = \,\,\,1.6 \pm 0.4\,\, \% , 
\nonumber \\ 
&& 
Br(D^{*0}\rightarrow D^0\pi^0)_{\rm exp}\,\, = 61.9 \pm 2.9 \,\, \%, \quad  \,
Br(D^{*0}\rightarrow D^0\gamma)_{\rm exp}\,\, \,\,\,\, = 38.1 \pm 2.9\,\, \% 
\label{eq:branching-fractions}
\end{eqnarray}
%%%%%%%%%%%%%%%%%%%%%%%%%%%%%%%%%%%%%%%%%%%%%%%%%%%%%%%%%%%%%%%%%%%%%%%%
which have been compiled by the Particle Data Group~\cite{PDG14}, we estimate 
rates for exclusive decays of $D^{*+}$ as  
%%%%%%%%%%%%%%%%%%%%%%%%%%%%%%%%%%%%%%%%%%%%%%%%%%%%%%%%%%%%%%%%%%%%%%%%
\begin{eqnarray}
&& 
\Gamma(D^{*+}\rightarrow D^0\pi^+)_{\rm exp} = 56.5 \pm 1.3\,\, {\rm keV}, \quad
\Gamma(D^{*+}\rightarrow D^+\pi^0)_{\rm exp} = 25.6 \pm 0.7\,\, {\rm keV}, %\quad  
\nonumber \\  
&& 
\Gamma(D^{*+}\rightarrow D^+\gamma)_{\rm exp}\,\, = \,\,\,\, 1.3 \pm 0.4\,\,\, 
{\rm keV}. 
                                                                    \label{eq:rates-of-charged-D^*-exp}
\end{eqnarray}
%%%%%%%%%%%%%%%%%%%%%%%%%%%%%%%%%%%%%%%%%%%%%%%%%%%%%%%%%%%%%%%%%%%%%%%%
However, exclusive decay rates of $D^{*0}$ are not known, because its full-width is not 
determined yet. 
Therefore, we estimate them under the isospin $SU_I(2)$ symmetry, below. 
Rate for $D^*\rightarrow D\pi$ decay is given by 
%%%%%%%%%%%%%%%%%%%%%%%%%%%%%%%%%%%%%%%%%%%%%%%%%%%%%%%%%%%%%%%%%%%%%%%%
\begin{equation}
\Gamma(D^*\rightarrow D\pi) 
= \frac{|\bm{k}_\pi|}{24\pi m_{D^{*}}^2}\sum_{pol}|M(D^*\rightarrow D\pi)|^2,
                                                                      \label{eq:formula-of-decay-rate}
\end{equation}
%%%%%%%%%%%%%%%%%%%%%%%%%%%%%%%%%%%%%%%%%%%%%%%%%%%%%%%%%%%%%%%%%%%%%%%%
where $\bm{k}_\pi$ is the center-of-mass (c.m.) momentum of pion in the final state,   
and the amplitude is written as 
%%%%%%%%%%%%%%%%%%%%%%%%%%%%%%%%%%%%%%%%%%%%%%%%%%%%%%%%%%%%%%%%%%%%%%%%
\begin{equation}
M(D^*(p)\rightarrow D(p')\pi(k)) =  g_{D^*\bar{D}\pi}e_\mu(p)q^\mu, \quad (q = p' - k)
                                                                                 \label{eq:Feynman-amp}
\end{equation}
%%%%%%%%%%%%%%%%%%%%%%%%%%%%%%%%%%%%%%%%%%%%%%%%%%%%%%%%%%%%%%%%%%%%%%%%
in this note. 
Here, $g_{D^*\bar{D}\pi}$ and  $e_\mu(p)$ are the $D^*\bar{D}\pi$ coupling strength 
and the polarization vector of $D^*$, respectively. 
By using a hard pion technique  in the infinite momentum frame 
(IMF)~\cite{Oneda-Terasaki-suppl}, 
i.e., by taking $\bm{k}\rightarrow 0$ in the $\bm{p}\,(\parallel z$-axis) 
$\rightarrow \infty$ frame, the amplitude is approximated as  
%%%%%%%%%%%%%%%%%%%%%%%%%%%%%%%%%%%%%%%%%%%%%%%%%%%%%%%%%%%%%%%%%%%%%%%%
\begin{eqnarray}
&&
\lim _{\bm{p}\rightarrow\infty, \bm{k}\rightarrow 0} M
(D^*\rightarrow D\pi) =  
\Bigl(\frac{m_{D^*}^2 - m_D^2}{f_\pi}\Bigr)\langle{D|A_{\bar{\pi}}|D^*}\rangle 
\nonumber\\
&&\hspace{37mm}
=  - g_{D^*\bar{D}\pi}\Bigl(\frac{m_{D^*}^2 - m_{D}^2}{m_{D^*}}\Bigr),
                                                                               \label{eq:hard-pion-approx}
\end{eqnarray}
%%%%%%%%%%%%%%%%%%%%%%%%%%%%%%%%%%%%%%%%%%%%%%%%%%%%%%%%%%%%%%%%%%%%%%%%
under the partially conserved axial-vector current (PCAC) hypothesis~\cite{PCAC}, 
where $A_{\pi}$ and $f_\pi$ are the axial-charge with the flavor of pion and the pion 
decay constant, respectively. 
Thus, we get 
%%%%%%%%%%%%%%%%%%%%%%%%%%%%%%%%%%%%%%%%%%%%%%%%%%%%%%%%%%%%%%%%%%%%%%%%
\begin{equation}
g_{D^*\bar{D}\pi} = - \Bigl(\frac{m_{D^*}}{f_\pi}\Bigr)\langle{D|A_{\bar{\pi}}|D^*}\rangle,  
                                                                                \label{eq:hybrid-rel}
\end{equation}
%%%%%%%%%%%%%%%%%%%%%%%%%%%%%%%%%%%%%%%%%%%%%%%%%%%%%%%%%%%%%%%%%%%%%%%%
which is considered as a meson version of the Goldberger-Treiman relation~\cite{GT}. 
We here assume that asymptotic matrix elements of $A_{\pi}$ (matrix elements of 
$A_{\pi}$ taken between single hadron states with the infinite momentum), 
$\langle{D|A_{\bar{\pi}}|D^*}\rangle$'s, satisfy the $SU_I(2)$ symmetry, i.e., 
%%%%%%%%%%%%%%%%%%%%%%%%%%%%%%%%%%%%%%%%%%%%%%%%%%%%%%%%%%%%%%%%%%%%%%%%
$\langle{D^+|A_{{\pi^+}}|D^{*0}}\rangle = {2}\langle{D^0|A_{{\pi^0}}|D^{*0}}\rangle 
= -{2}\langle{D^+|A_{{\pi^0}}|D^{*+}}\rangle = \langle{D^0|A_{{\pi^-}}|D^{*+}}\rangle$. 
%%%%%%%%%%%%%%%%%%%%%%%%%%%%%%%%%%%%%%%%%%%%%%%%%%%%%%%%%%%%%%%%%%%%%%%%
Under this approximation, we obtain 
%%%%%%%%%%%%%%%%%%%%%%%%%%%%%%%%%%%%%%%%%%%%%%%%%%%%%%%%%%%%%%%%%%%%%%%%
\begin{eqnarray}
&&  
\left.\begin{tabular}{l}
$\Gamma(D^{*+}\rightarrow D^+\pi^0)_{SU_I(2)} 
= 0.463\Gamma(D^{*+}\rightarrow D^0\pi^+) = 26.1 \pm 0.6 \,\,{\rm keV}$, 
\vspace{1mm}\\            %\nonumber \\    && 
$\Gamma(D^{*0}\rightarrow D^0\pi^0)_{\rm SU_I(2)} \,\,\,
= 0.580\Gamma(D^{*+}\rightarrow D^0\pi^+) = 37.1 \pm 1.0 \,\,{\rm keV}$, 
\end{tabular}\right. 
                                                                  \label{eq:rates-of-D^{*}-in-SU(2)}
\end{eqnarray}
%%%%%%%%%%%%%%%%%%%%%%%%%%%%%%%%%%%%%%%%%%%%%%%%%%%%%%%%%%%%%%%%%%%%%%%%
where we have used $f_{\pi^0} = f_{\pi^\pm}/\sqrt{2}$ and inserted the measured 
$\Gamma(D^{*+}\rightarrow D^0\pi^+)_{\rm exp} = 56.5 \pm 1.3\,\, {\rm keV}$ in 
Eq.~(\ref{eq:rates-of-charged-D^*-exp}) into $\Gamma(D^{*+}\rightarrow D^0\pi^+)$ 
in Eq.~(\ref{eq:rates-of-D^{*}-in-SU(2)}). 
The estimated $\Gamma(D^{*+}\rightarrow D^+\pi^0)_{SU_I(2)}$ is consistent with 
the measured rate in Eq.~(\ref{eq:rates-of-charged-D^*-exp}). 
This implies that the asymptotic $SU_I(2)$ symmetry ($SU_I(2)$ symmetry in 
asymptotic matrix elements) works well in pion emitting stromg decays of $D^*$  
mesons. 
The estimated $\Gamma(D^{*0}\rightarrow D^0\pi^0)_{SU_I(2)}$ in 
Eq.~(\ref{eq:rates-of-D^{*}-in-SU(2)}) and the measured branching 
fraction $Br(D^{*0}\rightarrow D^0\pi^0)_{\rm exp}$ in Eq.~(\ref{eq:branching-fractions}) 
lead to the full-width $(\Gamma_{D^{*0}})_{SU_I(2)}  = 59.9 \pm 3.3 \,\,{\rm keV}$ of 
$D^{*0}$. 
%%%%%%%%%%%%%%%%%%%%%%%%%%%%%%%%%%%%%%%%%%%%%%%%%%%%%%%%%%%%%%%%%%%%%%%%
Using the above $(\Gamma_{D^{*0}})_{SU_I(2)}$ and the measured branching fraction 
$Br(D^{*0}\rightarrow D^0\gamma)_{\rm exp}$ in Eq.~(\ref{eq:branching-fractions}), 
we obtain 
%%%%%%%%%%%%%%%%%%%%%%%%%%%%%%%%%%%%%%%%%%%%%%%%%%%%%%%%%%%%%%%%%%%%%%%%
\begin{equation}
\Gamma(D^{*0}\rightarrow D^0\gamma)_{SU_I(2)}  = 22.8 \pm 2.2 \,\, {\rm keV}. 
                                                                    \label{eq:rates-of-D^{*0}-in-SU(2)}
\end{equation}
%%%%%%%%%%%%%%%%%%%%%%%%%%%%%%%%%%%%%%%%%%%%%%%%%%%%%%%%%%%%%%%%%%%%%%%%
For later convenience, we here list the the following ratio of rates which is obtained 
from Eqs.~(\ref{eq:rates-of-charged-D^*-exp}{) and (\ref{eq:rates-of-D^{*0}-in-SU(2)}), 
%%%%%%%%%%%%%%%%%%%%%%%%%%%%%%%%%%%%%%%%%%%%%%%%%%%%%%%%%%%%%%%%%%%%%%%%
\begin{eqnarray}
&& 
\frac{\Gamma(D^{*0}\rightarrow D^0\gamma)_{SU_I(2)}}
          {\Gamma(D^{*+}\rightarrow D^+\gamma)_{\rm exp}} = 17.5\pm 5.7.
                                                                           \label{eq:ratio-of-rates-SU(2)}
\end{eqnarray}
%%%%%%%%%%%%%%%%%%%%%%%%%%%%%%%%%%%%%%%%%%%%%%%%%%%%%%%%%%%%%%%%%%%%%%%%

Next, we calculate rates for radiative decays of $D^*$ mesons. 
They are given in the form  
%%%%%%%%%%%%%%%%%%%%%%%%%%%%%%%%%%%%%%%%%%%%%%%%%%%%%%%%%%%%%%%%%%%%%%%%
\begin{equation}
\Gamma(D^*\rightarrow D\gamma) 
= \frac{|\bm{k}_\gamma|}{24\pi m_{D^{*}}^2}\sum_{pol}|M(D^*\rightarrow D\gamma)|^2,
                                                               \label{eq:formula-of-rate for-rad-decay}
\end{equation}
%%%%%%%%%%%%%%%%%%%%%%%%%%%%%%%%%%%%%%%%%%%%%%%%%%%%%%%%%%%%%%%%%%%%%%%%
where $\bm{k}_\gamma$ is the c.m. momentum of $\gamma$ in the final state. 
We here factor out the polarization independent part $A(D^*\rightarrow D\gamma)$ 
of the amplitude $M(D^*\rightarrow D\gamma)$ and write it in the form, 
%%%%%%%%%%%%%%%%%%%%%%%%%%%%%%%%%%%%%%%%%%%%%%%%%%%%%%%%%%%%%%%%%%%%%%%%
\begin{equation}
A(D^*\rightarrow D\gamma) 
= \sum_{V = \rho^0,\omega,\phi,\psi}
\Bigl[\frac{X_{V}(k_\gamma^2=0)}{m_V^2}\Bigr]g_{D^*\bar{D}V},
                                                                           \label{eq:truncated-amplitude}
\end{equation}
%%%%%%%%%%%%%%%%%%%%%%%%%%%%%%%%%%%%%%%%%%%%%%%%%%%%%%%%%%%%%%%%%%%%%%%%
under the vector meson dominance hypothesis (VMD)~\cite{VMD}, where 
$X_V(k_\gamma^2=0)$'s denote the photon-vector meson 
($V = \rho^0,\,\omega,\,\phi$ and $J/\psi$) coupling strengths on the photon mass 
shell. ($J/\psi$ will be written as $\psi$ hereafter.) 
In this manner, the amplitudes for the $D^{*+(0)}\rightarrow D^{+(0)}\gamma$ and 
$D_s^{*+}\rightarrow D_s^+\gamma$ decays are explicitly given by  
%%%%%%%%%%%%%%%%%%%%%%%%%%%%%%%%%%%%%%%%%%%%%%%%%%%%%%%%%%%%%%%%%%%%%%%%
\begin{eqnarray}
&& 
A(D^{*+}\rightarrow D^+\gamma) \,
= g_{D^{*+}D^-\rho^0}\frac{X_\rho(0)}{m_\rho^2} \,
+ g_{D^{*+}D^-\omega}\frac{X_\omega(0)}{m_\omega^2} 
+ g_{D^{*+}D^-\psi}\frac{X_\psi(0)}{m_\psi^2}, \\
&&
A(D^{*0}\rightarrow D^0\gamma) \,\,\,
= g_{D^{*0}\bar{D}^0\rho^0}\frac{X_\rho(0)}{m_\rho^2} \,\,\,
+ g_{D^{*0}\bar{D}^0\omega}\frac{X_\omega(0)}{m_\omega^2} \,\,
+ g_{D^{*0}\bar{D}^0\psi}\frac{X_\psi(0)}{m_\psi^2}, \\
&&
A(D_s^{*+}\rightarrow D_s^+\gamma) 
= g_{D_s^{*+}D_s^-\phi}\frac{X_\phi(0)}{m_\phi^2} \,\,
+ g_{D_s^{*+}D_s^-\psi}\frac{X_\psi(0)}{m_\psi^2}, 
\end{eqnarray}
%%%%%%%%%%%%%%%%%%%%%%%%%%%%%%%%%%%%%%%%%%%%%%%%%%%%%%%%%%%%%%%%%%%%%%%%
where it has been assumed that the $\omega$-$\phi$-$\psi$ mixing is ideal and 
the $D_{(s)}^*\bar{D}_{(s)}V$ vertices satisfy the OZI rule~\cite{OZI}. 

To study numerically rates for the above radiative decays, we need to know values of 
$X_V(0)$'s. 
They can be estimated from values of $\gamma V$ transition moments ($\gamma_V$'s) 
which are obtained from analyses in atomic number ($A$) dependence of forward 
cross sections of photoproductions of vector mesons on various targets. 
The results have been compiled as 
%%%%%%%%%%%%%%%%%%%%%%%%%%%%%%%%%%%%%%%%%%%%%%%%%%%%%%%%%%%%%%%%%%%%%%%%
\begin{eqnarray}
&& 
X_{\rho}(0)\, =\,\,\,\,0.033\pm 0.003 \,\,({\rm GeV})^2, \quad\,
X_\omega(0) = 0.011\pm 0.001 \,\,({\rm GeV})^2, 
\nonumber\\
&&
X_{\phi}(0) = -0.018\pm 0.004 \,\,({\rm GeV})^2. \quad
                                                                           \label{eq:photon-V-couplings}
\end{eqnarray}
%%%%%%%%%%%%%%%%%%%%%%%%%%%%%%%%%%%%%%%%%%%%%%%%%%%%%%%%%%%%%%%%%%%%%%%%
in \cite{VMD-KT} from data on $\gamma_V$'s given in the references quoted therein.   
However, we have updated the value of $X_\psi(0)$ as 
%%%%%%%%%%%%%%%%%%%%%%%%%%%%%%%%%%%%%%%%%%%%%%%%%%%%%%%%%%%%%%%%%%%%%%%%
\begin{equation}
X_\psi(0) = 0.15 \pm 0.02\,\,({\rm GeV})^2              \label{eq:photon-psi-couplings}
\end{equation}
%%%%%%%%%%%%%%%%%%%%%%%%%%%%%%%%%%%%%%%%%%%%%%%%%%%%%%%%%%%%%%%%%%%%%%%%
by using the measured rates for the $\psi\rightarrow\eta_c\gamma$ and 
$\eta_c\rightarrow\gamma\gamma$ decays, because data on forward cross section of 
$\psi$ photoproduction seem to be still unstable~\cite{Chudakov}. 
Next, it should be recalled~\cite{KT-ff} that a measure of the flavor symmetry breaking 
in hadronic interactions is given by the form factor $f_+(0)$ of related vector-current 
matrix element at zero momentum transfer squared. 
Its values have been compiled as~\cite{PDG96},  
%%%%%%%%%%%%%%%%%%%%%%%%%%%%%%%%%%%%%%%%%%%%%%%%%%%%%%%%%%%%%%%%%%%%%%%%
$f_+^{(\pi K)}(0) = 0.961\pm 0.008$,  %\cite{Leutwyler}, 
$f_+^{(\bar{K} D)}(0) = 0.74\pm 0.03$,  %\cite{PDG96}, 
$f_+^{(\pi D)}(0)/f_+^{(\bar{K} D)}(0) = 1.00\pm 0.13$ (FNAL-E687) 
and  $0.99\pm 0.08$ (CLEO). 
%%%%%%%%%%%%%%%%%%%%%%%%%%%%%%%%%%%%%%%%%%%%%%%%%%%%%%%%%%%%%%%%%%%%%%%%
These results suggest that the $SU_f(3)$ symmetry works well (even in the 
open-charm world), while the $SU_f(4)$ is broken to an extent of $20 - 30$ per cent. 
Therefore, it is assumed that the $D_{(s)}^*\bar{D}_{(s)}V$ coupling strengths satisfy 
the flavor $SU_f(3)$ symmetry, 
%%%%%%%%%%%%%%%%%%%%%%%%%%%%%%%%%%%%%%%%%%%%%%%%%%%%%%%%%%%%%%%%%%%%%%%%
\begin{equation}
\sqrt{2}g_{D^{*0}\bar{D}^0\omega} = \sqrt{2}g_{D^{*+}{D}^-\omega} 
= \sqrt{2}g_{D^{*0}\bar{D}^0\rho^0} = -\sqrt{2}g_{D^{*+}{D}^-\rho^0} 
= g_{D_s^{*+}{D}_s^-\phi},                                                 \label{eq:SU(3)-symmetry}
\end{equation}
%%%%%%%%%%%%%%%%%%%%%%%%%%%%%%%%%%%%%%%%%%%%%%%%%%%%%%%%%%%%%%%%%%%%%%%%
while deviation of $D_{(s)}^{*}\bar{D}_{(s)}\psi$ couplings from their $SU_f(4)$ symmetry 
limit is parameterized by  
%%%%%%%%%%%%%%%%%%%%%%%%%%%%%%%%%%%%%%%%%%%%%%%%%%%%%%%%%%%%%%%%%%%%%%%%
\begin{equation}
x = \frac{g_{D^{*+}{D}^-\psi}}{\sqrt{2}g_{D^{*+}{D}^-\omega} }
= \frac{g_{D^{*0}\bar{D}^0\psi}}{\sqrt{2}g_{D^{*0}\bar{D}^0\omega} }
= \frac{g_{D_s^{*+}{D}_s^-\psi}}{g_{D_s^{*+}{D_s^-}\phi} },
                                                   \label{eq:parametrization-of-SU(4)-symm-break}
\end{equation}
%%%%%%%%%%%%%%%%%%%%%%%%%%%%%%%%%%%%%%%%%%%%%%%%%%%%%%%%%%%%%%%%%%%%%%%%
where $x = 1$ in the $SU_f(4)$ symmetry limit. In this way, we can give ratios of rates 
for the above radiative decays by %
%%%%%%%%%%%%%%%%%%%%%%%%%%%%%%%%%%%%%%%%%%%%%%%%%%%%%%%%%%%%%%%%%%%%%%%%
\begin{center} 
\begin{table}[t]       
\begin{quote}
%\caption{
Table~I. 
Ratios of rates for radiative decays of open-charm vector mesons. 
The parameter $x$ is defined in the text. 
\end{quote} \vspace{2mm}
%%%%%%%%%%%%%%%%%%%%%%%%%%%%%%%%%%%%%%%%%%%%%%%%%%%%%%%%%%%%%%%%%%%%%%%%
\begin{tabular}{|c|c|c|c|c|c|c|}
\hline
$x$ & 1.0 & 0.9 & 0.8 & 0.7 & 0.6 & 0.5 
\\
\hline
$\displaystyle{\frac{\Gamma(D^{*0}\rightarrow D^0\gamma)}
       {\Gamma(D^{*+}\rightarrow D^+\gamma)}}$ & 42.4  & 30.6 & 22.9 & 17.5 
& 13.7 & 10.8  
\\
\hline
$\displaystyle{\frac{\Gamma(D_s^{*+}\rightarrow D_s^+\gamma)}
       {\Gamma(D^{*+}\rightarrow D^+\gamma)}}$ & 0.0276 & 0.0780 & 0.134  
& 0.189  & 0.241 & 0.289   
\\
\hline
\end{tabular}
%\vspace{3mm}\\
%%%%%%%%%%%%%%%%%%%%%%%%%%%%%%%%%%%%%%%%%%%%%%%%%%%%%%%%%%%%%%%%%%%%%%%%
\end{table}\vspace{-4mm}
\end{center}
%%%%%%%%%%%%%%%%%%%%%%%%%%%%%%%%%%%%%%%%%%%%%%%%%%%%%%%%%%%%%%%%%%%%%%%%
the unknown parameter $x$. 
Their numerical results are listed in Table~I.  
As seen in the table, the ratio of rates in Eq.~(\ref{eq:ratio-of-rates-SU(2)}) can be 
reproduced for $0.8\gtrsim x \gtrsim 0.6$, as expected from the above discussions. 
When $x = 0.7$ is taken, our ratio 
%%%%%%%%%%%%%%%%%%%%%%%%%%%%%%%%%%%%%%%%%%%%%%%%%%%%%%%%%%%%%%%%%%%%%%%%
${\Gamma(D^{*0}\rightarrow D^0\gamma)_{x=0.7}}
/{\Gamma(D^{*+}\rightarrow D^+\gamma)_{x=0.7}} = 17.5$ 
%%%%%%%%%%%%%%%%%%%%%%%%%%%%%%%%%%%%%%%%%%%%%%%%%%%%%%%%%%%%%%%%%%%%%%%%
agrees to the central value of the phenomenological ratio in 
Eq.~(\ref{eq:ratio-of-rates-SU(2)}), and then 
%%%%%%%%%%%%%%%%%%%%%%%%%%%%%%%%%%%%%%%%%%%%%%%%%%%%%%%%%%%%%%%%%%%%%%%%
\begin{equation}
\frac{\Gamma(D_s^{*+}\rightarrow D_s^+\gamma)}
          {\Gamma(D^{*+}\rightarrow D^+\gamma)}\Bigr|_{x=0.7} = 0.189  
                                                                                    \label{eq:ratio-D_s-to-D}
\end{equation}
%%%%%%%%%%%%%%%%%%%%%%%%%%%%%%%%%%%%%%%%%%%%%%%%%%%%%%%%%%%%%%%%%%%%%%%%
is obtained. 
Insertion of $\Gamma(D^{*+}\rightarrow D^+\gamma)_{\rm exp} = 1.3\pm 0.4$ keV in 
Eq.~(\ref{eq:rates-of-charged-D^*-exp}) into the denominator of 
Eq.~(\ref{eq:ratio-D_s-to-D}) leads to 
%%%%%%%%%%%%%%%%%%%%%%%%%%%%%%%%%%%%%%%%%%%%%%%%%%%%%%%%%%%%%%%%%%%%%%%%
\begin{equation}
\Gamma(D_s^{*+}\rightarrow D_s^+\gamma)_{x=0.7} = (0.25\pm 0.08)\,\,{\rm keV}. 
                                                                             \label{eq:rate-rad-decay-D_s }
\end{equation}
%%%%%%%%%%%%%%%%%%%%%%%%%%%%%%%%%%%%%%%%%%%%%%%%%%%%%%%%%%%%%%%%%%%%%%%%
From this result, we can estimate the rate for the isospin non-conserving 
$D_s^{*+}\rightarrow D_s^+\pi^0$ decay. 
Because the measured branching fraction has been given by 
$Br(D_s^{*+}\rightarrow D_s^+\gamma)_{\rm exp} = 94.2\pm 0.7$ per cent, 
the full-width of $D_s^{*+}$ is estimated as 
$(\Gamma_{D_s^{*+}})_{x=0.7} = 0.27\pm 0.09$ keV. 
Therefore, the rate for the isospin non-conserving $D_s^{*+}\rightarrow D_s^+\pi^0$ 
decay is estimated as 
%%%%%%%%%%%%%%%%%%%%%%%%%%%%%%%%%%%%%%%%%%%%%%%%%%%%%%%%%%%%%%%%%%%%%%%%
\begin{equation}
\Gamma(D_s^{*+}\rightarrow D_s^+\pi^0)_{x=0.7} 
= 0.016\pm 0.006\,\, {\rm keV}, 
                                                                    \label{eq:rate-for-I-non-cons-decay}
\end{equation}
%%%%%%%%%%%%%%%%%%%%%%%%%%%%%%%%%%%%%%%%%%%%%%%%%%%%%%%%%%%%%%%%%%%%%%%%
because of $Br(D_s^{*+}\rightarrow D_s^+\pi^0)_{\rm exp} = 5.8\pm 0.7$ per cent. 
%%%%%%%%%%%%%%%%%%%%%%%%%%%%%%%%%%%%%%%%%%%%%%%%%%%%%%%%%%%%%%%%%%%%%%%%
The rates in Eqs.~(\ref{eq:rate-rad-decay-D_s }) and 
(\ref{eq:rate-for-I-non-cons-decay}) provide  
%%%%%%%%%%%%%%%%%%%%%%%%%%%%%%%%%%%%%%%%%%%%%%%%%%%%%%%%%%%%%%%%%%%%%%%%
\begin{equation}
\frac{\Gamma(D_s^{*+}\rightarrow D_s^+\pi^0)_{x=0.7}}
 {\Gamma(D_s^{*+}\rightarrow D_s^+\gamma)_{x=0.7}} = 0.064\pm 0.028. 
                                                         \label{eq:ratio-of-rates-for-pi-to-gamma}
\end{equation}
%%%%%%%%%%%%%%%%%%%%%%%%%%%%%%%%%%%%%%%%%%%%%%%%%%%%%%%%%%%%%%%%%%%%%%%%
This result is consistent with the measured ratio, $0.62\pm 0.007$~\cite{PDG14}, 
though our result contains large uncertainties. 
%%%%%%%%%%%%%%%%%%%%%%%%%%%%%%%%%%%%%%%%%%%%%%%%%%%%%%%%%%%%%%%%%%%%%%%%
This implies that the result in Eq.~(\ref{eq:rate-for-I-non-cons-decay}) is natural, and  
therefore, it is compared with our rate 
$\Gamma(D_s^{*+}\rightarrow D_s^+\pi^0)_{\eta\pi^0}$ 
for the isospin non-conserving decay through the $\eta\pi^0$ mixing, below. 

The charm strange vector meson $D_s^{*+}$ has no kinematically-allowed hadronic 
isospin-consreving decay. 
Its kinematically-allowed decay $D_s^{*+}\rightarrow D_s^+\pi^0$ is isospin 
non-conserving and its rate is given by  
%%%%%%%%%%%%%%%%%%%%%%%%%%%%%%%%%%%%%%%%%%%%%%%%%%%%%%%%%%%%%%%%%%%%%%%%
\begin{equation}
\Gamma(D_s^{*+}\rightarrow D_s^+\pi^0) 
= \frac{|\bm{k}_{\pi^0}|^3}{6\pi m_{D_s^{*+}}^2}\Bigl|\Bigl(\frac{m_{D_s^{*+}}}{f_{\pi^0}}
\Bigr)\langle{D_s^+|A_{\pi^0}|D_s^{*+}}\rangle\Bigr|^2
                                                          \label{eq:rate-for-D_s^*}
\end{equation}
%%%%%%%%%%%%%%%%%%%%%%%%%%%%%%%%%%%%%%%%%%%%%%%%%%%%%%%%%%%%%%%%%%%%%%%%
in the same way as the $D^*\rightarrow D\pi$, where $\bf{k}_{\pi^0}$ is the c.m. 
momentum of $\pi^0$ in the final state and $f_{\pi^0} = f_{\pi^\pm}/\sqrt{2}$.  
We here assume that the decay proceeds through the $\eta\pi^0$ 
mixing~\cite{Cho-Wise} with the mixing parameter $\epsilon$. 
In this case, the asymptotic matrix element 
$\langle{D_s^+|A_{\pi^0}|D_s^{*+}}\rangle$ is given by 
%%%%%%%%%%%%%%%%%%%%%%%%%%%%%%%%%%%%%%%%%%%%%%%%%%%%%%%%%%%%%%%%%%%%%%%%
\begin{equation}
\langle{D_s^+|A_{\pi^0}|D_s^{*+}}\rangle = \epsilon\langle{D_s^+|A_{\eta}|D_s^{*+}}\rangle 
= -\epsilon\sin(\Theta)\langle{D_s^+|A_{\eta_s}|D_s^{*+}}\rangle
                                                                                    \label{eq:I-non-cons}
\end{equation}
%%%%%%%%%%%%%%%%%%%%%%%%%%%%%%%%%%%%%%%%%%%%%%%%%%%%%%%%%%%%%%%%%%%%%%%%
under the asymptotic $SU_f(3)$ symmetry~\cite{Oneda-Terasaki-suppl}, where 
$\Theta = \chi + \theta_P$ with the $\eta\eta'$ mixing angle $\theta_P$ and 
$\chi = \arccos(\sqrt{1/3}) = \arcsin(\sqrt{2/3}) = 54.7^\circ$, and $A_{\eta_s}$ is 
the axial-charge with the flavor of $\{\bar{s}s\}$ component of $\eta$. 
The asymptotic $SU_f(3)$ symmetry implies that the asymptotic matrix elements 
satisfy %~\cite{Hallock} 
%%%%%%%%%%%%%%%%%%%%%%%%%%%%%%%%%%%%%%%%%%%%%%%%%%%%%%%%%%%%%%%%%%%%%%%%
$\langle{D_s^+|A_{\eta_s}|D_s^{*+}}\rangle = \langle{D^0|A_{\pi^-}|D^{*+}}\rangle$, 
%%%%%%%%%%%%%%%%%%%%%%%%%%%%%%%%%%%%%%%%%%%%%%%%%%%%%%%%%%%%%%%%%%%%%%%%
so that the following ratio of rates is obtained, 
%%%%%%%%%%%%%%%%%%%%%%%%%%%%%%%%%%%%%%%%%%%%%%%%%%%%%%%%%%%%%%%%%%%%%%%%
\begin{eqnarray}
&&\hspace{-5mm}
R_{\eta\pi^0} = \frac{\Gamma(D_s^{*+}\rightarrow D_s^+\pi^0)_{\eta\pi^0}}
                            {\Gamma(D^{*+}\rightarrow D^0\pi^+)} 
= \Bigl|\frac{\bm{k}_{\pi^0}}{\bm{k}_{\pi^+}}\Bigr|^3
              \Bigl(\frac{f_\pi}{f_{\pi^0}}\Bigr)^2\bigl[\epsilon\sin(\Theta)\bigr]^2.
                                                              \label{eq:ratio-of-rates-pi^0-to-gamma}
\end{eqnarray}
%%%%%%%%%%%%%%%%%%%%%%%%%%%%%%%%%%%%%%%%%%%%%%%%%%%%%%%%%%%%%%%%%%%%%%%%
Because the $\eta\eta'$ mixing angle $\theta_P$ has not been determined yet, 
we consider the following three cases,  
(i) $\theta_P = -11.4^\circ$ (estimated by using the quadratic G-M-O mass 
formula~\cite{GMO}), 
(ii) $\theta_P = -24.5^\circ$ (estimated by using  the linear G-M-O mass formula) 
and (iii) $\theta_P = -14.1^\circ\pm 2.8^\circ$ (estimated by a lattice QCD 
simulation~\cite{lattice}), as listed in \cite{PDG14}.  
In each of these cases, the ratio in Eq.~(\ref{eq:ratio-of-rates-pi^0-to-gamma}) is 
given by 
%%%%%%%%%%%%%%%%%%%%%%%%%%%%%%%%%%%%%%%%%%%%%%%%%%%%%%%%%%%%%%%%%%%%%%%%
\begin{eqnarray}
&&\hspace{-5mm}
R_{\eta\pi^0} = 1.76|\epsilon|^2\,\, {\rm in\,\, (i)},\quad
0.946|\epsilon|^2\,\,  \,\,{\rm in\,\, (ii)},\quad 1.58|\epsilon|^2  \,\,{\rm in\,\,(iii)}, 
\end{eqnarray}
%%%%%%%%%%%%%%%%%%%%%%%%%%%%%%%%%%%%%%%%%%%%%%%%%%%%%%%%%%%%%%%%%%%%%%%%
where about 20 per cent errors of the estimated $\theta_P$ in (iii) have been 
neglected.  
Although the mixing parameter $\epsilon$ was given, long time ago, as 
$\epsilon = 0.0105 \pm 0.0013$~\cite{Dalitz} which is $O(\alpha)$ as expected,  
it is now drastically improved.  
When we take $\epsilon = 0.01058$ and 
$\Gamma(D^{*+}\rightarrow D^0\pi^+)_{\rm exp} = 56.5\pm 1.3$ keV 
in Eq.~(\ref{eq:rates-of-charged-D^*-exp}) as the input data, we obtain the results 
listed in Table II. 
Comparing them with Eq.~(\ref{eq:rate-for-I-non-cons-decay}), we find that the cases 
(i) and (iii) are  favored, while the result in (ii) seems to be not favored, though our 
results involve large uncertainties arising from the $SU_f(4)$ symmetry breaking 
parameter $x$. 
%%%%%%%%%%%%%%%%%%%%%%%%%%%%%%%%%%%%%%%%%%%%%%%%%%%%%%%%%%%%%%%%%%%%%%%%
\begin{center} 
\begin{table}[t]       
\begin{quote}
%\caption{
Table~II. 
Rate %$\Gamma(D_s^{*+}\rightarrow D_s^+\pi^0)_{\eta\pi^0}$ 
for the isospin 
non-conserving $D_s^{*+}\rightarrow D_s^+\pi^0$ decay through the $\eta\pi^0$ 
mixing. 
In (i), (ii) and (iii), the $\eta\eta'$ mixing angle is taken as 
$\theta_P = -11.4^\circ$, $\theta_P = -24.5^\circ$ and $\theta_P = -14.1^\circ$, 
respectively. 
The results should be compared with 
$\Gamma(D_s^{*+}\rightarrow D_s^+\pi^0)_{x = 0.7} =  0.016\pm 0.006$  keV 
in Eq.~(\ref{eq:rate-for-I-non-cons-decay}). 
\end{quote} \vspace{2mm}
%%%%%%%%%%%%%%%%%%%%%%%%%%%%%%%%%%%%%%%%%%%%%%%%%%%%%%%%%%%%%%%%%%%%%%%%
\begin{tabular}{|l|c|c|c|}
\hline
& (i) $\theta_P = -11.4^\circ$ & (ii) $\theta_P = -24.5^\circ$ 
& (iii)  $\theta_P = -14.1^\circ$
\\
\hline
$\Gamma(D_s^{*+}\rightarrow D_s^+\pi^0)_{\eta\pi^0}$ & $ 0.0111\pm 0.0003$ keV 
& $0.0060\pm 0.0002$ keV & $0.0100\pm 0.0003$ keV
\\
\hline
\end{tabular}
%\vspace{1mm}\\
%%%%%%%%%%%%%%%%%%%%%%%%%%%%%%%%%%%%%%%%%%%%%%%%%%%%%%%%%%%%%%%%%%%%%%%%
\end{table} %\vspace{-4mm}
\end{center}
%%%%%%%%%%%%%%%%%%%%%%%%%%%%%%%%%%%%%%%%%%%%%%%%%%%%%%%%%%%%%%%%%%%%%%%%

We here compare our results on decays of charmed vector mesons with those from 
a hybrid model~\cite{Bardeen}.  %in which %as one of existing studies.  
In this model, pion emitting decays are calculated in a chiral Lagrangian approach in 
which $\eta$ participating in the $\eta\pi^0$ mixing is assumed to be of a pure 
$SU_f(3)$-octet (without any $\eta\eta'$ mixing) and radiative decays are studied in 
a constituent quark model. 
Regarding with the $SU_I(2)$ symmetry, this model predicted  
%%%%%%%%%%%%%%%%%%%%%%%%%%%%%%%%%%%%%%%%%%%%%%%%%%%%%%%%%%%%%%%%%%%%%%%%
$\Gamma(D^{*+}\rightarrow D^0\pi^+)/\Gamma(D^{*+}\rightarrow D^+\pi^0)
\simeq 2$, 
%%%%%%%%%%%%%%%%%%%%%%%%%%%%%%%%%%%%%%%%%%%%%%%%%%%%%%%%%%%%%%%%%%%%%%%%
in consistency with our result and experiments~\cite{PDG14} in 
Eq.~(\ref{eq:rates-of-charged-D^*-exp}). 
The ratio of rates  
%%%%%%%%%%%%%%%%%%%%%%%%%%%%%%%%%%%%%%%%%%%%%%%%%%%%%%%%%%%%%%%%%%%%%%%%
$\Gamma(D^{*+}\rightarrow D^+\gamma)/\Gamma(D^{*0}\rightarrow D^0\gamma) 
= 0.058\pm 0.015$ 
%%%%%%%%%%%%%%%%%%%%%%%%%%%%%%%%%%%%%%%%%%%%%%%%%%%%%%%%%%%%%%%%%%%%%%%%
which was predicted by the hybrid model is accidentally consistent with our estimate in 
Eq.~(\ref{eq:ratio-of-rates-SU(2)}). 
As for the isospin non-conserving $D_s^{*+}\rightarrow D_s^+\pi^0$ decay, this theory  
has predicted %its rate as 
%%%%%%%%%%%%%%%%%%%%%%%%%%%%%%%%%%%%%%%%%%%%%%%%%%%%%%%%%%%%%%%%%%%%%%%%
%\begin{equation}
$\Gamma(D_s^{*+}\rightarrow D_s^+\pi^0)_{\rm hyb} = 0.0079\,\,{\rm keV}$, 
%\end{equation}
%%%%%%%%%%%%%%%%%%%%%%%%%%%%%%%%%%%%%%%%%%%%%%%%%%%%%%%%%%%%%%%%%%%%%%%%
by assuming that the decay is caused by the $\eta\pi^0$ mixing. 
Although the value of the mixing parameter is close to ours, the above result is 
approximately a half of our semi-phenomenological estimate  
Eq.~(\ref{eq:rate-for-I-non-cons-decay}) with which our results in (i) and (iii) are 
compatible.  
Regarding with the radiative decay, the hybrid model has predicted 
%%%%%%%%%%%%%%%%%%%%%%%%%%%%%%%%%%%%%%%%%%%%%%%%%%%%%%%%%%%%%%%%%%%%%%%%
%\begin{equation}
$\Gamma(D_s^{*+}\rightarrow D_s^+\gamma)_{\rm hyb} = 0.43\,\,{\rm keV}$ 
%\end{equation}
%%%%%%%%%%%%%%%%%%%%%%%%%%%%%%%%%%%%%%%%%%%%%%%%%%%%%%%%%%%%%%%%%%%%%%%%
by using the constituent quark model. 
However, this result is larger by about 70 per cent than our  estimate 
Eq.~(\ref{eq:rate-rad-decay-D_s }) under the VMD. 
In cosequence, the ratio of rates has been predicted as  
%%%%%%%%%%%%%%%%%%%%%%%%%%%%%%%%%%%%%%%%%%%%%%%%%%%%%%%%%%%%%%%%%%%%%%%%
\begin{equation}
\Gamma(D_s^{*+}\rightarrow D_s^+\pi^0)/\Gamma(D_s^{*+}\rightarrow D_s^+\gamma)
= 0.018   
\end{equation}
%%%%%%%%%%%%%%%%%%%%%%%%%%%%%%%%%%%%%%%%%%%%%%%%%%%%%%%%%%%%%%%%%%%%%%%%
which is far from the measured one $0.062\pm 0.008$~\cite{PDG14}. 
On the other hand, our results on the rate 
$\Gamma(D_s^{*+}\rightarrow D_s^+\pi^0)_{\eta\pi^0}$ 
in (i) and (iii) reproduce approximately Eq.~(\ref{eq:rate-for-I-non-cons-decay})
as seen in Table~II, though the same $\eta\pi^0$ mixing is assumed as the origin of 
the isospin non-conservation in the $D_s^{*+}\rightarrow D_s^+\pi^0$ decay, and  
our estimates of the ratio of rates 
$\Gamma(D_s^{*+}\rightarrow D_s^+\pi^0)_{\eta\pi^0}
    /\Gamma(D_s^{*+}\rightarrow D_s^+\gamma)_{x = 0.7}
= 0.044\pm 0.015$ in (i) and $0.040 \pm 0.013$ in (iii) 
are compatible with the measurement, though their errors %of our estimates 
are still large. 

In summary, we have studied decay property of charmed vector mesons, applying 
a hard pion technique in the IMF (an innovation of the well-known soft pion technique) 
to pion emitting decays and the VMD to radiative decays.  
As the result, we have seen that the $SU_I(2)$ symmetry works well in 
$D^*\rightarrow D\pi$ decays and predicted the full-width of $D^{*0}$ under the 
$SU_I(2)$ symmetry. 
Therefore, measurements of full-width of $D^{*0}$ are awaited. 
Then, radiative decays of $D^*$ and $D_s^{*+}$ have been studied under the VMD, by 
assuming that the $D_{(s)}^{*}\bar{D}_{(s)}V,\,\,(V = \rho,\,\,\omega,\,\,\phi)$ coupling 
strengths satisfy the $SU_f(3)$ symmetry, while the deviation from the $SU_f(4)$ 
symmetry limit of the $D_{(s)}^{*}\bar{D}_{(s)}\psi$ couplings has been considered 
phenomenologically and has been taken to be 30 percent (i.e., $x = 0.7$). 
The isospin non-conserving $D_s^{*+}\rightarrow D_s^+\pi^0$ decay has been studied 
by assuming that it proceeds through the $\eta\pi^0$ mixing. 
Its rate is explicitly dependent on the $\eta\eta'$ mixing angle $\theta_P$. 
When $\theta_P = -11.4^\circ$ (from the quadratic G-M-O mass formula) and 
$\theta_P = -14.1^\circ$ (from a lattice QCD simulation) have been taken, our 
values of ratio of rates 
%%%%%%%%%%%%%%%%%%%%%%%%%%%%%%%%%%%%%%%%%%%%%%%%%%%%%%%%%%%%%%%%%%%%%%%
$\Gamma(D_s^{*+}\rightarrow D_s^+\pi^0)_{\eta\pi^0}
  /\Gamma(D_s^{*+}\rightarrow D_s^+\gamma)_{x=0.7}$ 
%%%%%%%%%%%%%%%%%%%%%%%%%%%%%%%%%%%%%%%%%%%%%%%%%%%%%%%%%%%%%%%%%%%%%%%
have been compatible with the measured one, while the result with 
$\theta_P = -24.5^\circ$ (from the linear G-M-O mass formula) has not been favored 
by the measured ratio. 
Therefore, determinations of the $\eta\eta'$ mixing angle will be important to 
establish the role of the $\eta\pi^0$ mixing as the origin of the isospin 
non-conservation in the $D_s^{*+}\rightarrow D_s^+\pi^0$ decay.  

%%%%%%%%%%%%%%%%%%%%%%%%%%%%%%%%%%%%%%%%%%%%%%%%%%%%%%%%%%%%%%%%%%%%%%%
\section*{Acknowledgments}    
The author would like to thank Professor T.~Hyodo for valuable discussions and 
comments. 
He also would like to appreciate Professor H.~Kunitomo for careful reading of the 
manuscript. 
%%%%%%%%%%%%%%%%%%%%%%%%%%%%%%%%%%%%%%%%%%%%%%%%%%%%%%%%%%%%%%%%%%%%%%%

%%%%%%%%%%%%%%%%%%%%%%%%%%%%%%%%%

%\end{references}
%%%%%%%%%%%%%%%%%%%%%%%%%%%%%%%%%%%%%%%%%%%%%%
\end{document}